\documentclass[11pt]{article}
\usepackage{moriond,epsfig}
\pdfoutput=1

\def\Journal#1#2#3#4{{#1} {\bf #2}, #3 (#4)}

\def\be{\begin{equation}}
\def\ee{\end{equation}}
\def\bea{\begin{eqnarray}}
\def\eea{\end{eqnarray}}

\begin{document}
\vspace*{4cm}
\title{Dilepton mass edge measurement in SUSY events with CMS}

\author{ N. Mohr }

\address{I. Physikalisches Institut B, RWTH Aachen University, Otto-Blumenthal-Strasse,\\
D-52074 Aachen, Germany}

\maketitle\abstracts{
Within the mSUGRA model, the observability of the decay of the next to lightest neutralino $\tilde{\chi}_2^0$ into leptons and the lightest neutralino $\tilde{\chi}_1^0$ has been studied using a full simulation of the CMS detector. The final state signature consists of two opposite sign leptons, several hard jets and missing transverse energy. The expected precision of the measurement of the dilepton mass edge  is reported for 1~fb$^{-1}$ of data, including systematic and statistic uncertainties, comparing two benchmark points with different signatures.}

\section{Introduction}

The standard model of particle physics (SM) provides no solution for pressing questions arising from astrophysical observations such as dark matter. In Supersymmetry (SUSY) a natural candidate for dark matter can be found if R-parity conservation is assumed. Supersymmetric particles (sparticles) have not been observed up to now what implies that they have to be heavy. On the other hand to provide a solution for the hierarchy problem their masses have to be in the TeV range.

The long anticipated start of the Large Hadron Collider (LHC) in 2009 will allow to explore this new TeV range. With its center of mass energy of 10 TeV in 2010 it will allow to probe supersymmetric models very early on. A key point after discovery will be the determination of the sparticle properties. Because the lightest neutralino escapes detection, no mass peaks can be observed in SUSY decay chains. Of special interest are robust signatures such as mass edges in leptonic final states which can be probed with the CMS experiment~\cite{cms}.

\section{Leptonic decay of the next to lightest neutralino}

\begin{table}[t]
\caption{mSUGRA benchmark points LM1 and LM9.\label{tab:Bench}}
\vspace{0.4cm}
\begin{center}
\begin{tabular}{|l|c|c|c|c|c|c|} \hline
    &$m_0$ [GeV]    &	$m_{1/2}$ [GeV]	&	$A_0$ [GeV]	&	$\tan{\beta}$	&	sign $\mu$ & $m_{ll,max}$ [GeV]	\\ \hline
LM1 &60		    &	250		&	0		&	10		&	+1	&  78,2	\\ \hline
LM9 &1450	    &	175	        &	0		&	50		&	+1	&  62,9	\\ \hline
\end{tabular}
\end{center}
\end{table}

The leptonic decay of the next to lightest neutralino provides a clear signature of two opposite sign same flavour leptons which allows to trigger and identify the events in the hadron collider environment at LHC. In the framework of minimal Supergravity (mSUGRA) two leptonic decay modes of the next to lightest neutralino are possible a 2-body or 3-body decay. Two benchmark points (called LM1 and LM9) have been studied which correspond to these decay modes. The mSUGRA parameters and the endpoint in the dilepton invariant mass of both points are listed in Tab.~\ref{tab:Bench}. The low energy mass spectra of the two benchmark points have been calculated using the Softsusy code~\cite{soft}. Depending on the mass spectrum different decay modes are possible.

\begin{figure}
\includegraphics[angle=90,width=0.49\textwidth]{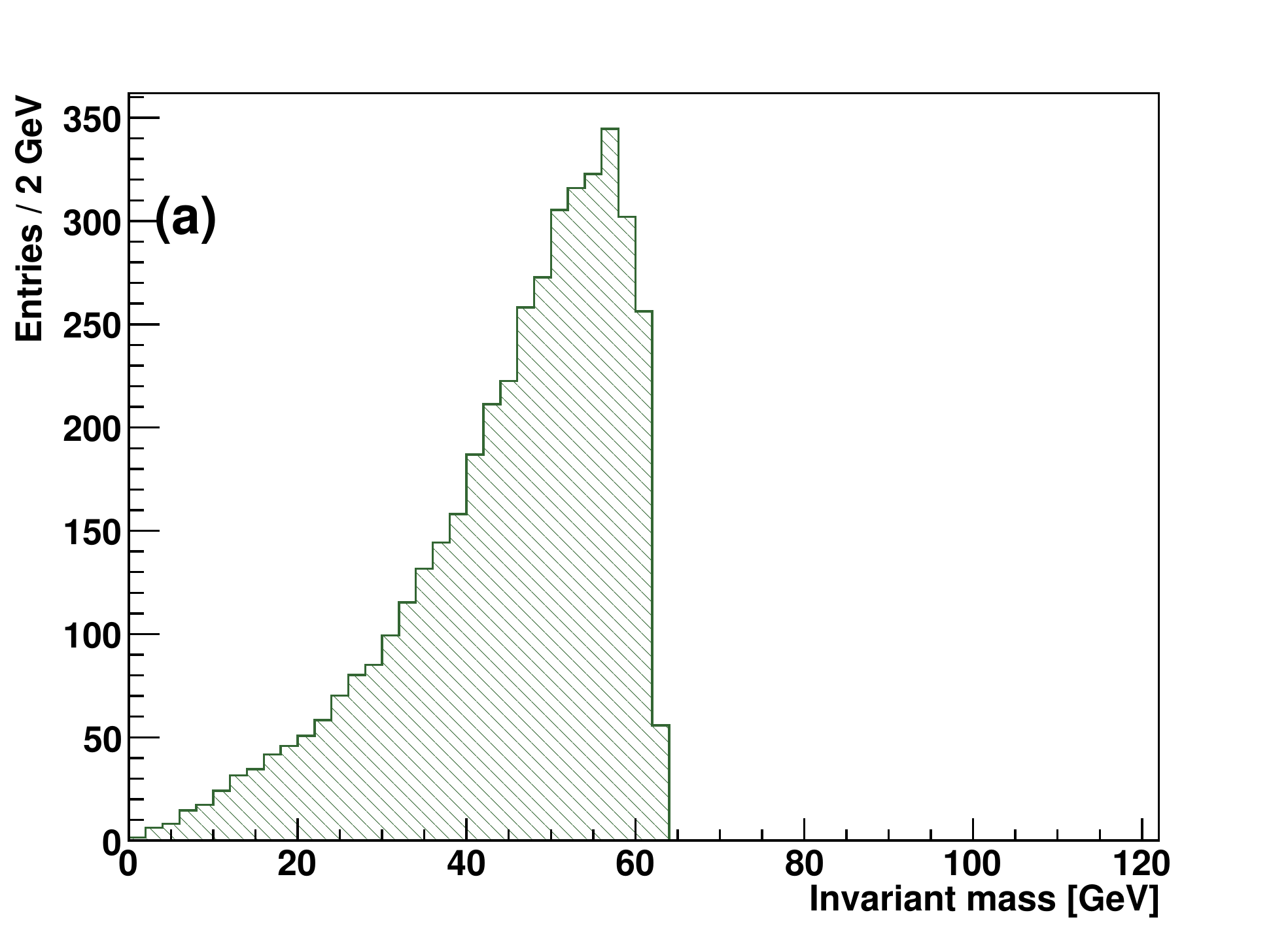}
\includegraphics[angle=90,width=0.49\textwidth]{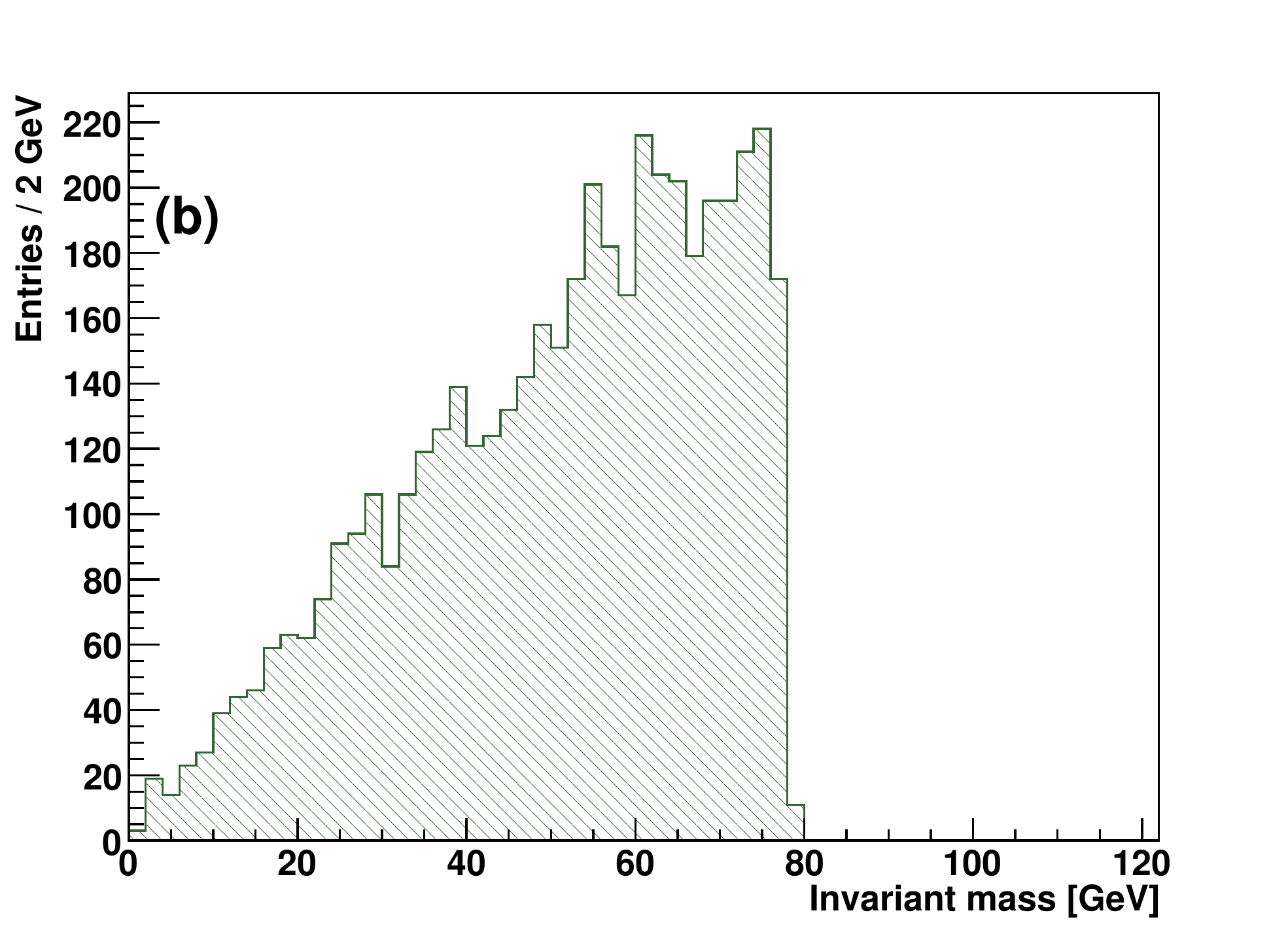}
\caption{Invariant mass distribution of leptons from the decay of the next to lightest neutralino are shown at generator level. The 3-body decay occurs at LM9 (a) and the 2-body decay at LM1 (b).
\label{fig:decaymodes}}
\end{figure}

A mass difference of the neutralinos smaller than the Z-boson mass and any slepton mass leads to a three body decay. In that case the endpoint represents directly the mass difference of the two lightest neutralinos
\be
m_{ll,max} = m_{\tilde{\chi}_2^0} -m_{\tilde{\chi}_1^0}.
\label{eq:neutralino2body}
\ee
The shape of the distribution depends on the mSUGRA parameters and is shown in Fig.~\ref{fig:decaymodes}~(a) for the LM9 benchmark point.

A two body decay occurs via a real slepton and is allowed if one slepton is lighter than the mass difference of the neutralinos. In that case the endpoint can be expressed by
\be
\left( m_{ll}^{max} \right)^2 = \frac{\left( m_{\tilde{l}}^2 -m_{\tilde{\chi}_2^0}^2 \right) \left( m_{\tilde{\chi}_2^0}^2 -m_{\tilde{\chi}_2^0}^2\right)}{m_{\tilde{l}}^2}
\label{eq:neutralino2body}
\ee
The shape of the mass edge results only from kinematics and is triangular as shown in Fig.~\ref{fig:decaymodes}~(b).

\section{Event selection}

In order to select the signal events a single leptonic trigger has been used. Because of the long cascade decays the final state consist of a high number of hard jets. The escaping neutralino leads to missing transverse energy in the detector. The selection requires three jets with $E_T^{j1} > 120$~GeV,  $E_T^{j2} > 100$~GeV and $E_T^{j3} > 80$~GeV which are corrected for response. Additionally a missing transverse energy measured in the calorimeter (corrected for muons and jet energy) of at least 125 GeV is required. Finally two well identified and isolated ($\sum p_T < 1.5$~GeV for tracks within a cone of $\Delta R<0.3$) leptons with $p_T^{l} > 10$~GeV are required.


As main Standard Model backgrounds events with real isolated leptons are considered. To simulate these events the full CMS detector simulation has been used. The processes $t\bar{t}$+jets, $Z$+jets and $W$+jets have been simulated using the Alpgen MC~\cite{alpgen}. The QCD background, which does not contain isolated high $p_T$ leptons has been simulated with the Pythia code~\cite{pythia}. The SUSY sample is simulated using the Softsusy, SUSYHit~\cite{susyhit} and Pythia codes.

\begin{figure}
\includegraphics[angle=90,width=0.49\textwidth]{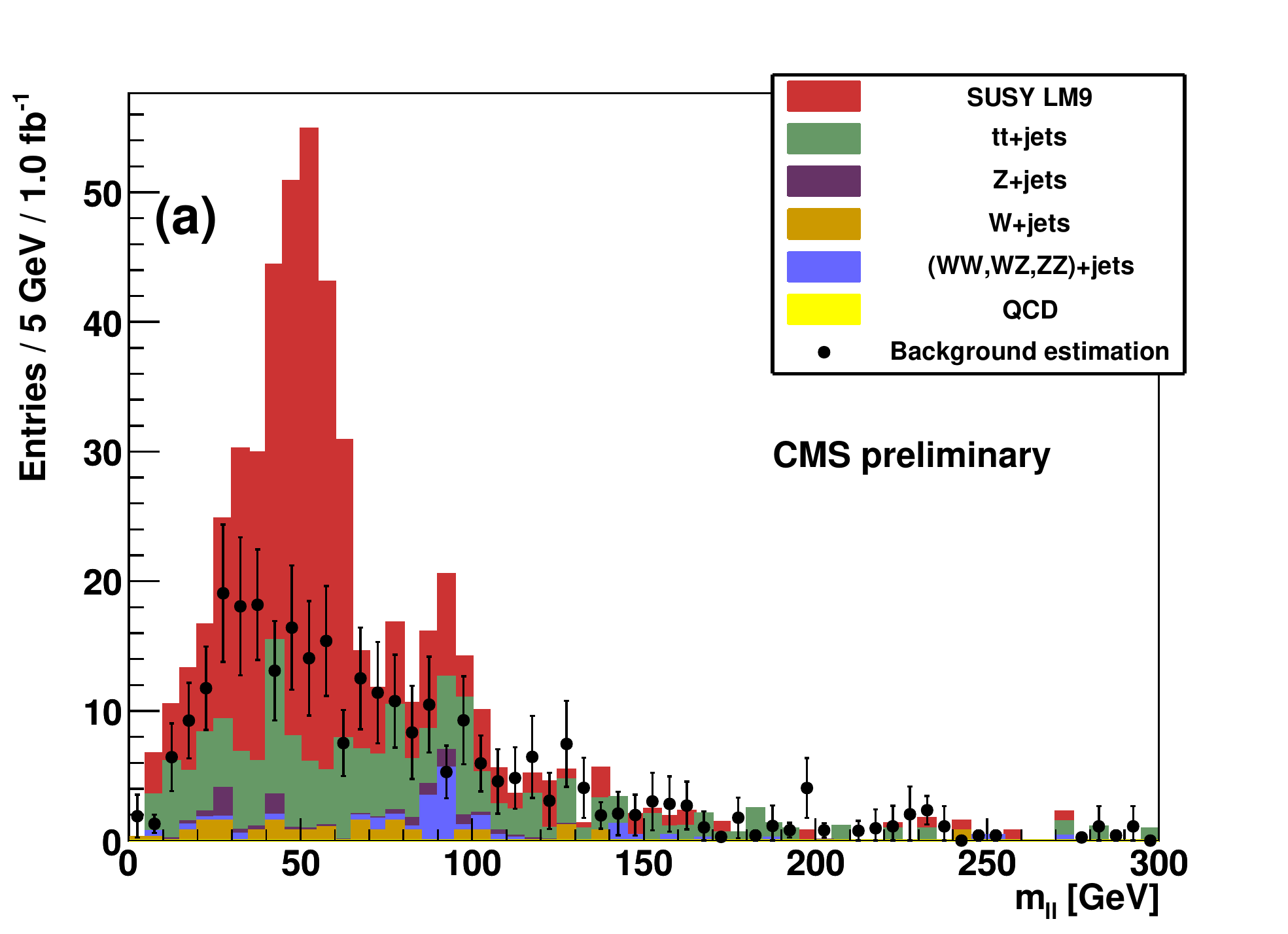}
\includegraphics[angle=90,width=0.49\textwidth]{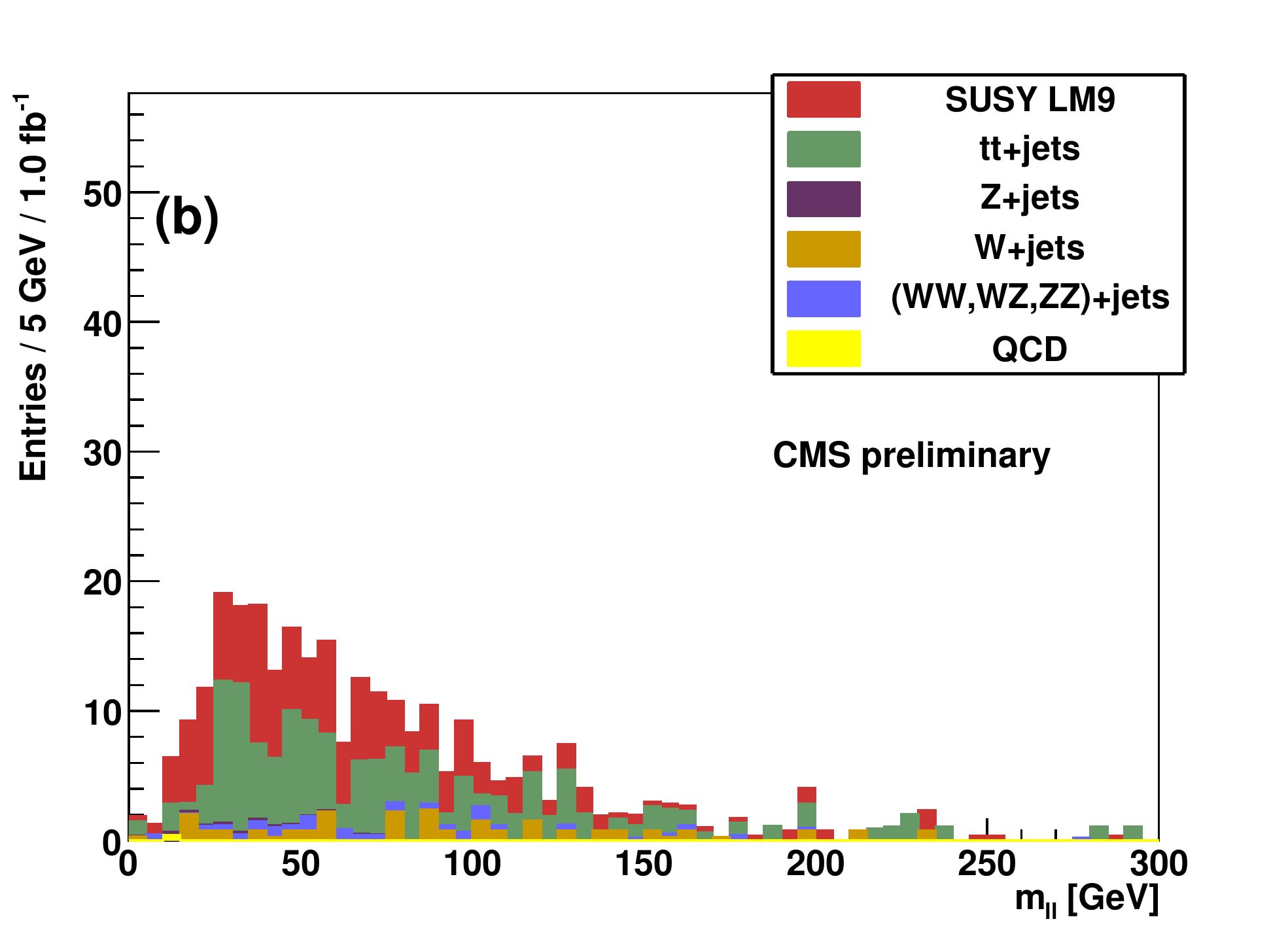}
\caption{The invariant mass distribution for same flavour opposite sign electron and muon pairs is shown (a). The opposite flavour opposite sign flavour background (b) consists mainly of top pair events and is used to measure the flavour symmetric background. The points in (a) represent the background extrapolation from (b).
\label{fig:eemu}}
\end{figure}

If one applies the full selection on the simulated set of events the main remaining background consists of $t\bar{t}$+jets events Fig.~\ref{fig:eemu} (a). Every background where the two leptons are produced uncorrelated, e.g. $t\bar{t}$ events, can be measured directly from data by selection of opposite sign opposite flavour lepton pairs as shown in Fig.~\ref{fig:eemu} (b). Using these wrong pairings one can predict the background distribution in the opposite sign same flavour lepton distribution. The method relies only on the knowledge of the lepton reconstruction efficiency which can be measured from data using events with a Z-bosons decaying leptonically~\cite{electron}.

\section{Measurement of the mass edge}

\begin{figure}
\includegraphics[angle=90,width=0.49\textwidth]{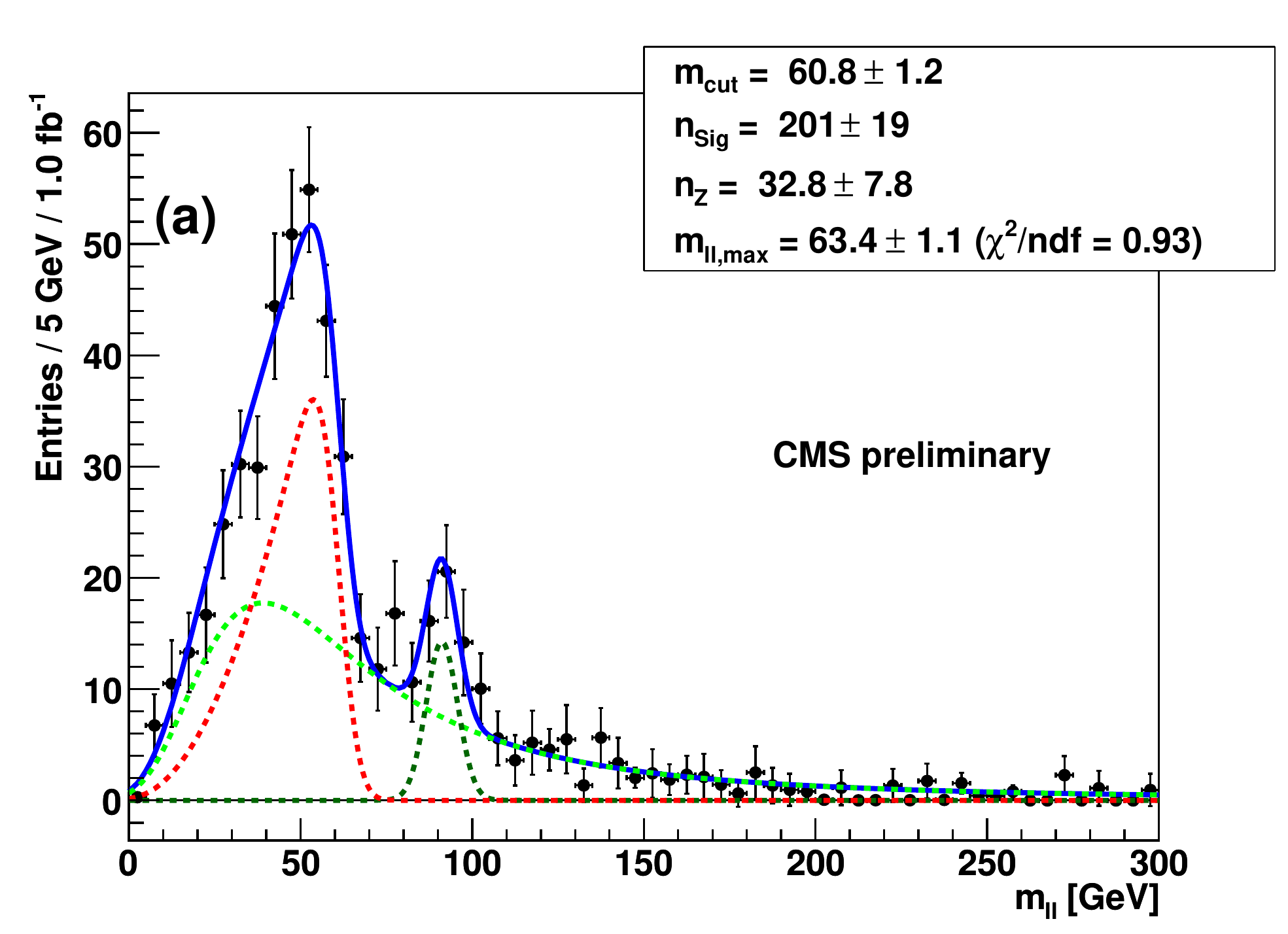}
\includegraphics[width=0.49\textwidth]{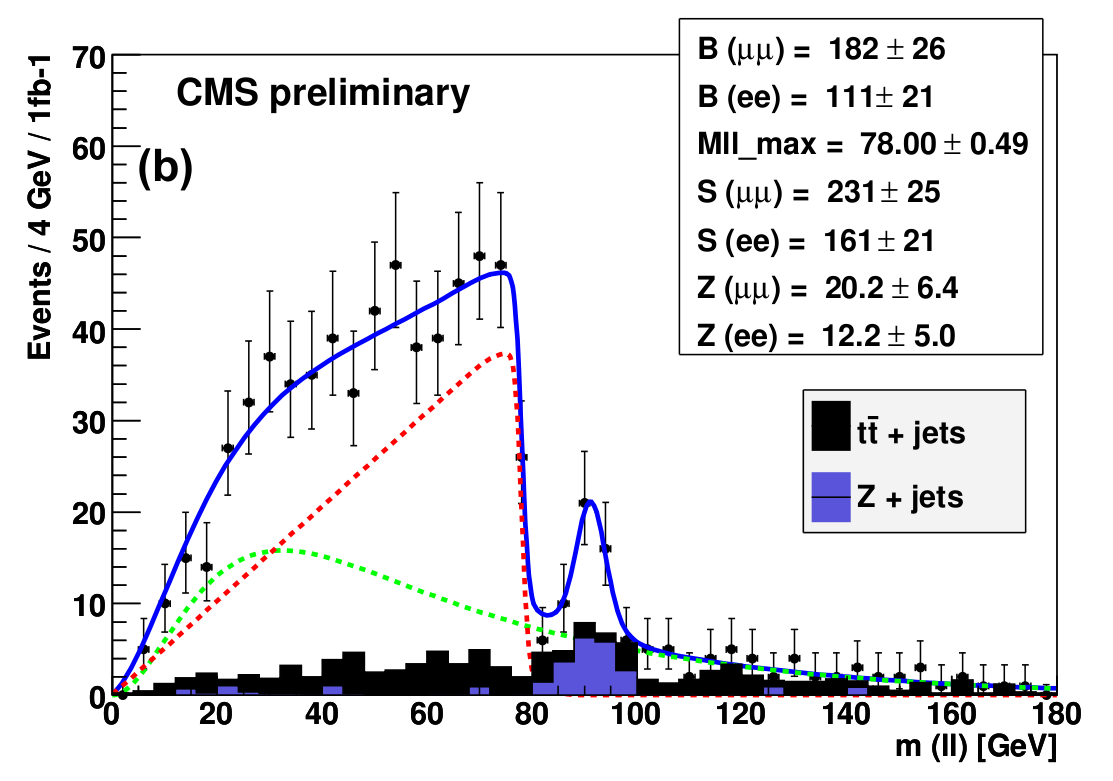}
\caption{Fit of the invariant mass distributions at LM9 (a) and LM1 (b). The dashed green line in (a) is the background landau function, the dark green curve a Breit-Wigner function (Z contribution) and dashed red line the SUSY model. The dashed green curve in (b) is the background model and the red dashed line the SUSY model (triangle). The points represent the monte carlo events and the blue line is the final fit.
\label{fig:fitlm9t175}}
\end{figure}

A combined fit consisting of three parts has been used to measure the mass edge. In case of a 3-body decay the signal model consists of a quadratic term convoluted with a gaussian
\be
f(x) = \frac{1}{\sqrt{2\pi}\sigma} \int\limits_{0}^{m_{cut}} dy \cdot y^2 e^{\frac{-\left( x- y \right)^2}{2\sigma^2}}.
\ee
As background model a Landau curve has been fitted to the flavour symmetric background distribution Fig.~\ref{fig:eemu}~(b). The Z-peak is modelled by a Breit-Wigner function convoluted with a gaussian. The estimator of the mass difference ($m_{cut}$) has been calibrated using MC templates with different values of the neutralino mass difference. The final fit for LM9 is shown in Fig.~\ref{fig:fitlm9t175}~(a) and yields a value of
\be
m_{ll,max} = \left( 63.4 \pm 1.1_{stat.} \pm 0.8_{syst.} \right) \textnormal{ GeV.}
\label{eq:mllLM1}
\ee

Similarly the endpoint in case of a two body decay has been studied~\cite{georgia}. In that case the theoretical shape is known and a triangle, which is not calibrated using MC templates, is used to fit the distribution. The final fit shown in Fig.~\ref{fig:fitlm9t175}~(b) for LM1 yields a value of 
\be
m_{ll,max} = \left( 78.0 \pm 0.6_{stat.} \pm 0.4_{syst.} \right) \textnormal{ GeV.}
\label{eq:mllLM1}
\ee

The systematic error is calculated taking into account a misalignment and miscalibration of the CMS detector as expected after $100$~pb$^{-1}$ of data. Additionally the uncertainties on the jet energy scale (10\%), the lepton energy scale and the acceptance have been included as well as uncertainties due to the fit model. It is found that this measurement is still dominated by statistical uncertainties at an integrated luminosity of 1~fb$^{-1}$. The difference in the shape of both distributions can be used to identify the type of decay mode on a goodness of fit basis.

\section{Conclusion}

The leptonic endpoint can be reconstructed within the first LHC data (1~fb$^{-1}$) if a low mass SUSY scenario is realised in nature. At both studied benchmark points the expected endpoint can be reproduced. The total error on the measurement is expected to be less than 3\%.

\section*{Acknowledgments}
Thanks to the members in the CMS SUSY group. This study is supported by the German Federal Ministry of Education and Research trough FSP~102.

\end{document}